\documentstyle[12pt]{article}
\newlength{\extraspace}
\newlength{\extraspaces}
\catcode`\@=11
\def\numberbysection{\@addtoreset{equation}{section}
\def\theequation{\arabic{section}.\arabic{equation}}}
\newcommand{\be}{\begin{equation}
\addtolength{\abovedisplayskip}{\extraspaces}
\addtolength{\belowdisplayskip}{\extraspaces}
\addtolength{\abovedisplayshortskip}{\extraspace}
\addtolength{\belowdisplayshortskip}{\extraspace}}
\newcommand{\ee}{\end{equation}}
\newcommand{\ba}{\begin{eqnarray}
\addtolength{\abovedisplayskip}{\extraspaces}
\addtolength{\belowdisplayskip}{\extraspaces}
\addtolength{\abovedisplayshortskip}{\extraspace}
\addtolength{\belowdisplayshortskip}{\extraspace}}
\newcommand{\ea}{\end{eqnarray}}
\newcommand{\newsection}[1]{
\vspace{7mm}
\pagebreak[3]
\addtocounter{section}{1}
\setcounter{subsection}{0}
\setcounter{footnote}{0}
\begin{center}
{\large {\bf \thesection. #1}}
\end{center}
\nopagebreak
\medskip
\nopagebreak
\hspace{3mm}}
\newcommand{\nonu}{\nonumber \\[.5mm]}
\newcommand{\A}{&\!\!\!}

\newcommand{\e}{\, {\rm e}}

\begin{document}
\addtolength{\baselineskip}{.7mm}
\thispagestyle{empty}
\begin{flushright}
OCHA--PP--153 \\
{\tt hep-th/0004179} \\ 
April, 2000
\end{flushright}
\vspace{25mm}
\begin{center}
{\Large{\bf Conformal Supergravity from the AdS/CFT Correspondence}} 
\\[20mm] 
{\sc Madoka Nishimura}
\footnote{\tt e-mail: nishi@sokewpie.phys.ocha.ac.jp} 
\\[3mm]
{\it Department of Physics, Ochanomizu University \\
2-1-1, Otsuka, Bunkyo-ku, Tokyo 112-8610, Japan} 
\end{center}
\vspace{25mm}
\begin{center}
{\bf Abstract}
\end{center}
Construction of a five dimensional conformal supergravity 
($D=5$ CSG) 
is attempted by applying the AdS/CFT correspondence to 
the F(4) AdS supergravity in six dimensions.
As a first step, local transformation laws of 
$D=5$ CSG have been established, from which the Weyl weights of 
the various fields in $D=5$ can be predicted. 
\newpage
%%
%%%%% Section 1 %%%%%%%%%%%%%%%%%%%%%%%%%%%%%%%%%%%%%%%%%%%%
%
\newsection{Introduction}
The main usage of the AdS/CFT conjecture has been 
concentrated on the study of the strong coupling limit 
of conformal field theories on the boundary 
\cite{MAL}, \cite{GKP}, \cite{WITTEN}, \cite{AGMOO}, 
since there is a one-to-one correspondence between 
the AdS supergravity fields in the bulk 
and the operators belonging to the representation of 
the superconformal group on the boundary. 
\par
Our studies have, however, been focused on 
the local symmetries, in which the local symmetries 
in the bulk become those on the boundary 
by applying a certain gauge fixing condition 
to the AdS supergravity fields. 
We obtain multiplets of the conformal supergravity 
(CSG) on the boundary \cite{LT}, \cite{NT}, \cite{NT3}, 
by taking the near boundary limit of the AdS supergravity 
in the bulk. 
These CSG fields play the roles of sources 
to the corresponding operators of a certain SCFT. 
Bulk local symmetries have some relations 
to the boundary global symmetries for the SCFT, 
though they are not related directly. 
For example, in the case of three dimensional bulk, 
we can reproduce the central charge 
of two dimensional CFT,
which is the coefficient of the CSG local anomalies 
from the AdS supergravity Lagrangian \cite{NT2}.
\par
Let us summarize our general strategy briefly. 
Consider a $(p+2)$-dimensional AdS supergravity, and 
let $r$ denote the direction normal to the boundary. 
The AdS supergravity has local symmetries 
under the general coordinate, 
the local Lorentz, 
the supersymmetry and the gauge transformations, 
if the theory is gauged. 
Here we emphasize that there is neither Weyl nor super Weyl 
transformations in the bulk. 
\par
These symmetries are induced on the boundary from 
the general coordinate transformation in the $r$-direction 
and the supersymmetry transformation, respectively, and they form 
the superconformal transformations.
Together with other local symmetries such as $p+1$-dimensional 
general coordinate transformations, 
they become local symmetries of the CSG. 
\par
Here, the important point is the choice of 
gauge fixing condition in the bulk. 
Our method is essentially independent of the dimension 
of the AdS space and the type of the supergravity.
The difference of the dimensions imposes 
only the different conditions on fermion fields.
\par
In the previous papers, 
according to the original AdS/CFT conjecture, 
we treated the type IIB string/M-theoretical cases. 
In this paper, we will study the six dimensional 
massive type IIA AdS supergravity 
and attempt to derive the corresponding CSG from it. 
\par
In the analysis we have done previously, 
we understand that the one dimensionally reduced CSG local symmetries 
are obtained from the original AdS supergravity 
with the same gauge and supersymmetry. 
So we can expect that the six dimensional local symmetries
of the AdS supergravity induce 
those of five dimensional CSG on the boundary. 
Unfortunately, 
few odd dimensional CSGs have been known so far.
\par
For that reason, our trial means to construct 
a certain type of CSG from the AdS supergravity. 
Our challenge may seem to be brave, 
because we have a remarkable 
difference to overcome in the six dimensional case. 
We can not define the chirality 
nor impose the Weyl condition on the boundary, 
since we are considering the space-time of five dimensions. 
The situation is in contrast to the two dimensional 
and six dimensional cases. 
Therefore, 
we will introduce a new projective gamma matrix $\tilde{\Gamma}$ 
instead of the chirality operator and get over the difference.
\par
Our convention of the metric is 
$\eta_{AB}={\rm diag} (-1,1,\cdots,1)$, 
where $A,B,\cdots = 0,1, \cdots ,5$ are 
six dimensional local Lorentz indices. 
The $16\times 16$ gamma matrices $\Gamma^A$ 
are defined by 
\be
\{ \Gamma^A, \Gamma^B \} = 2 \eta^{AB}, 
\ee
and we define 
\be
\Gamma_7 = \Gamma^0 \cdots \Gamma^5. 
\ee
%
%%%%% Section 2 %%%%%%%%%%%%%%%%%%%%%%%%%%%%%%%%%%%%%%%%%%%%
%
\newsection{$D=6$, F(4) AdS supergravity}
Let us start with the relationship between the supergravity 
and D-brane configuration. 
Although $D=6$, F(4) AdS supergravity \cite{ROMANS} 
has been constructed in 1980's, 
it was taken notice again 
in the D4-D8 system where its brane solution has known 
to become the AdS${}_6$ metric \cite{BO}. 
The warped compactification of the type IIA massive 
supergravity gives the solution for the sphere 
times the six dimensional F(4) AdS supergravity 
\cite{YOUM}, \cite{CLP}. 
On the boundary, a dual field theory is coupled 
to the five dimensional CSG derived from the AdS${}_6$ 
supergravity, 
and the operators of the dual field theory have been 
already discussed \cite{FKPZ}. 
\par
Let us go back to the original supergravity story. 
F(4) is the supergroup whose bosonic subalgebra 
is SO(2,5) $\times$ SU(2). 
The field contents are 
a six dimensional vielvein, $e_M{}^A$, 
three gauge vectors $A_M^I$ of SU(2) gauge group, 
an antisymmetric tensor field $B_{MN}$, 
a scalar field $\phi$, 
four Rarita-Schwinger fields $\psi_{Mi}$, 
and four spin-${1 \over 2}$ fields $\chi_i$. 
Here $M,N,\cdots(=0,1,2,\cdots,5)$ denote the world indices, 
$I,J,\cdots(=1,2,3)$ the vector indices of the gauge group SU(2) 
and $i,j,\cdots(=1,2)$ the spinor indices of SU(2), 
respectively. 
The spinor fields satisfy 
the SU(2) symplectic Majorana condition.
\par
The covariant derivative for arbitrary spinor $\epsilon_i$ 
is defined as follows: 
\be
{\cal D}_M \epsilon_i \equiv 
(\partial_M + {1 \over 4}\omega_M{}^{AB} \Gamma_{AB}) 
\epsilon_i 
     -i {1 \over 2}g A_M^I (\sigma^I)_i{}^j \epsilon_j, 
\ee
where $g$ is the coupling constant, and $\sigma^I$ are the Pauli matrices.
The local transformations are 
the six dimensional general coordinate, 
the local Lorentz, 
the super and SU(2) gauge transformations. 
With these local symmetries, 
the Lagrangian of $D=6$ F(4) AdS supergravity \cite{ROMANS}
stands up to 4-fermi terms as
\ba
e{}^{-1} L \A = \A 
     - {1 \over 4}R 
     - i{1 \over 2} 
          {\bar{\psi}}^i_M \Gamma^{MNP}{\cal D}_N \psi_{Mi}
     - i{1 \over 2} 
          {\bar{\chi}}^i \Gamma^M {\cal D}_M \chi_i
     -{1 \over 2} ({\cal D}^M \phi)({\cal D}_M \phi) 
     \nonu \A \A 
     - {1 \over 4} \e^{-2\sqrt{1 \over 2}\phi} 
          \left(m^2 B_{MN} B^{MN}
               + F_{MN}^I F^{MNI} \right) 
     - {3 \over 4} \e^{4\sqrt{1 \over 2} \phi} 
          \partial_{[M} B_{NP]} \partial^{[M} B^{NP]} 
     \nonu \A \A
     - {1\over 8} e^{MNPQRS} B_{MN}
          \left(
          {1 \over 3} m^2 B_{PQ}B_{RS}+F^I_{PQ}F^I_{RS}
          \right)
     \nonu \A \A 
     + {1 \over 4\sqrt{2}} \e^{- \sqrt{1 \over 2} \phi}
          \bar{\psi}^{Pi}\Gamma_{[P}\Gamma^{MN}\Gamma_{Q]}
	  \left(
	  mB_{MN} \delta_i^j
	  -i\Gamma_7 F_{MN}^I (\sigma^I)_i{}^j
	  \right)\psi^Q_j
     \nonu \A \A 
     - i {1 \over 4\sqrt{2}} \e^{- \sqrt{1 \over 2} \phi}
          \bar{\psi}^{Pi}\Gamma^{MN}\Gamma_P
	  \left(
	  mB_{MN} \delta^i_j
	  +i\Gamma_7 F_{MN}^I (\sigma^I)_i{}^j
	  \right)\chi_j 
     \nonu \A \A
     + {1 \over 8\sqrt{2}} {\bar{\chi}}^i \Gamma^{MN}
          \left(
          mB_{MN} \delta^i_j
	  +i\Gamma_7 F_{MN}^I (\sigma^I)_i{}^j
          \right) \chi_j
     \nonu \A \A 
     - {1 \over \sqrt{2}}(\partial_N \phi) 
          \bar{\psi}^i_M \Gamma^N \Gamma^M \chi_i
     -i{1 \over 8} \e^{2\sqrt{1 \over 2}\phi} 
          \bar{\psi}^{Mi}\Gamma_{[P}\Gamma_7 \Gamma^{QRS}
	  \partial_{[Q} B_{RS]}\Gamma_{N]} \psi^N_i 
     \nonu \A \A
     +{1 \over 4} \e^{2\sqrt{1 \over 2}\phi}
          \bar{\psi}^{Pi} \Gamma_7 \Gamma^{QRS} 
	  \partial_{[Q} B_{RS]}\Gamma_P \chi_i
     +i{1 \over 8} \e^{2\sqrt{1 \over 2}\phi}
          \bar{\psi}^{Mi}\Gamma_{[P}\Gamma_7 \Gamma^{QRS}
	  \partial_{[Q} B_{RS]} \chi_i
     \nonu \A \A
     + {1 \over 4\sqrt{2}} \left( 
          g \e^{\sqrt{1 \over 2}\phi} + m \e^{-3 \sqrt{1 \over 2}\phi}
          \right) 
          {\bar{\psi}}^i_M \Gamma^{MN} \Gamma_7 \psi_{N i} 
     \nonu \A \A 
     -i {1 \over 4\sqrt{2}} \left( 
          g \e^{\sqrt{1 \over 2}\phi} -3 m \e^{-3 \sqrt{1 \over 2}\phi}
          \right)
          {\bar{\psi}}^i{}_M \Gamma^M \Gamma_7 \chi_i 
     \nonu \A \A 
     + {1 \over 8\sqrt{2}} \left( 
          g \e^{\sqrt{1 \over 2}\phi} -7 m \e^{-3 \sqrt{1 \over 2}\phi}
          \right)
          {\bar{\chi}}^i \Gamma_7 \chi_i 
\nonu \A \A 
     + {1 \over 8} \left( 
          g^2 \e^{2 \sqrt{1 \over 2}\phi} 
          + 4gm \e^{-2 \sqrt{1 \over 2}\phi} 
          - m^2 \e^{-6 \sqrt{1 \over 2}\phi}
     \right), 
\ea
where $m$ is the parameter, 
and we have introduced $e^{MNPQRS}$ by
\be
e_{MNPQRS} = \Gamma_{MNPQRS} \Gamma_7. 
\ee 
The SU(2) vector gauge field strength is defined by 
\be
F_{MN}^I \equiv 
2 \partial_{[M} A_{N]}^I + g \epsilon^{IJK} A_M{}^J A_N{}^K.
\ee
The local supertransformations are up to 2-fermi terms
\ba
\delta_Q e_M{}^A \A \equiv \A 
     -i {\bar{\psi}}^i_M \Gamma^A \epsilon_i
\nonu 
\delta_Q A_M^I \A \equiv \A  
     i {1 \over \sqrt{2}} \e^{\sqrt{1 \over 2}\phi} (\sigma^I)_i{}^j 
     \left(
          {\bar{\psi}}^i_M \Gamma_7 \epsilon_j
          +  {1 \over 2} {\bar{\chi}}^i \Gamma_M \Gamma_7 \epsilon_j
     \right), 
\nonu 
\delta_Q B_{MN} \A \equiv \A 
      - i \e^{-2 \sqrt{1 \over 2}\phi} 
      \left(
      {\bar{\psi}}^i_{[M} \Gamma_{N]} \Gamma_7 \epsilon_i
      - i {1 \over 2} {\bar{\chi}}^i \Gamma_M \Gamma_7 \epsilon_i
     \right), 
\nonu 
\delta_Q \phi \A \equiv \A {1 \over \sqrt{2}} {\bar{\chi}}^i \epsilon_i, 
\label{transfs6} \\
\delta_Q \psi_{Mi} \A \equiv \A {\cal D}_M \epsilon_i
     - i {1 \over 8 \sqrt{2}} \left( 
          g \e^{\sqrt{1 \over 2}\phi} + m \e^{-3 \sqrt{1 \over 2}\phi}
          \right) 
          \Gamma_M \Gamma_7 \epsilon_i 
     \nonu \A \A 
     + i {1 \over 8\sqrt{2}} \e^{-\sqrt{1 \over 2}\phi} 
          \left( \Gamma_M{}^{PQ} -6 \delta_M^P \Gamma^Q \right)
          \left(
               m B_{PQ} \delta_i^j 
               - i \Gamma_7 F_{PQ}{}^I (\sigma^I)_i{}^j 
          \right) \epsilon_j 
     \nonu \A \A 
     + {1 \over 8} \e^{2 \sqrt{1 \over 2}\phi} 
          \Gamma_7 \Gamma^{PQR} \partial_{[P} B_{QR]} \epsilon_i, 
\nonu 
\delta_Q \chi_i \A \equiv \A 
     -i \sqrt{1 \over 2} \Gamma^M (\partial_M \phi) \epsilon_i 
      + {1 \over 4\sqrt{2}} \left( 
          g \e^{\sqrt{1 \over 2}\phi} -3 m \e^{-3 \sqrt{1 \over 2}\phi}
          \right)  \Gamma_7 \epsilon_i
     \nonu \A \A 
     + i {1 \over 4} \e^{2 \sqrt{1 \over 2}\phi} 
          \Gamma_7 \Gamma^{PQR} \partial_{[P} B_{QR]} \epsilon_i 
     \nonu 
     \A \A 
     - {1 \over 4\sqrt{2}} \e^{- \sqrt{1 \over 2}\phi} 
          \Gamma^{PQ} \left( 
               m B_{PQ} \delta_i^j 
               -i \Gamma_7 F_{PQ}{}^I (\sigma^I)_i{}^j 
          \right) \epsilon_j.
\nonumber
\ea
Here $\epsilon_i(i=1, 2)$ 
are the parameters of supertransformations, which 
satisfy the symplectic Majorana condition.
Thus the theory is 
$N=(2,2)$ D=6 AdS supergravity with the gauge group SU(2) 
(having  16 supercharges).
Note that we count the number of supersymmetry 
by the number of supertransformation parameters. 
The commutator of two supersymmetry transformations satisfies
\be
[\delta_{Q_1}(\epsilon_1),\delta_{Q_2}(\epsilon_2)] 
= \delta_{\rm G}(\xi^M)
+\delta_{\rm L}(\Sigma^{AB})+\delta_{\rm {SU(2)}}(\Lambda^I), 
\ee
where the parameters of 
the general coordinate transformation $\xi^M$, 
the local Lorentz transformation $\Sigma^{AB}$ and 
the SU(2) gauge transformation $\Lambda^I$
are defined respectively as, 
\ba
\xi^M \A = \A 
-i {\bar{\epsilon}}^i_2 \Gamma^M \epsilon_{1i}, 
\nonu
\Sigma^{AB} \A = \A 
- \xi^M \omega_M{}^{AB} 
\nonu \A \A 
+ {1 \over 4 \sqrt{2}} \e^{-\sqrt{1 \over 2}\phi} 
     {\bar{\epsilon}}_2^i
     \left( 
     \Gamma^{AB}{}_{CD} + 6 \delta^{AB}_{CD} 
     \right)
     \left(
          m B_{PQ} \delta_i^j 
          -i \Gamma_7 F_{PQ}^I (\sigma^I)_i{}^j 
     \right) e^{PC} e^{QD} \epsilon_{1j}
\nonu \A \A 
-{1 \over 4} {\bar{\epsilon}}_2^i 
     \left[
     \e^{2\sqrt{1 \over 2}\phi} \left(
     \Gamma^{AB}{}_{CDE}+6\delta^A_C\delta^B_D\Gamma_E
     \right) \partial^{[P}B^{QR]} e^{PC} e^{QD} e_R{}^E \right.
     \nonu
     \A \A 
     \left.
     + {1 \over \sqrt{2}}\left( 
          g \e^{\sqrt{1 \over 2}\phi} + m \e^{-3 \sqrt{1 \over 2}\phi}
          \right) \Gamma^{AB} \right] \epsilon_{1i},
\\
\Lambda^I \A = \A \xi^M A_M{}^I 
     + i{1 \over \sqrt{2}}(\sigma^I)_i{}^j 
     {\bar{\epsilon}}^i_2 \Gamma_7 \epsilon_{1j}. 
\nonumber
\ea
\par
In order to calculate the $r$-dependence for the fields and 
the transformation parameters, 
we need to expand the fields 
around the vacuum expectation value of the scalar field $\phi$. 
There are two stationary points: 
\be 
\phi = {1 \over 2 \sqrt{2}} \ln {m \over g}, \qquad 
	{1 \over 2 \sqrt{2}} \ln {3m \over g}.
\ee 
We can find that the former is not supersymmetric 
whereas the latter preserves the supersymmetry 
as is known by solving the Killing spinor equation. 
We choose the stable point $\phi_0$ as the latter one
\be
\phi_0 = {1 \over 2 \sqrt{2}} \ln {3m \over g}. 
\ee
At first sight, since this point is the maximum point 
of the potential, 
the supersymmetric vacuum does not seem to be stable. 
However, 
according to the extended version 
of Breitenlohner-Freedman stability condition 
\cite{BF}, \cite{MT}, 
we see that it should be stable. 
\newpage
%
%
%%%%% Section 3 %%%%%%%%%%%%%%%%%%%%%%%%%%%%%%%%%%%%%%%%%%%%
%
\newsection{The Gauge Fixing Conditions 
and the Boundary Behaviors for the Fields} 
In the following sections, 
we will construct the conformal supergravity transformations
from the AdS supergravity. 
First, we introduce the gauge fixing conditions such that 
the direction $r$ is normal to the boundary, 
and denote the other directions of the space as 
$\mu,\nu \cdots = 0,1,\cdots,4$. 
The local Lorentz indices, 
$a,b, \cdots$ runs among $0,1,\cdots,4$. 
We choose the AdS${}_6$ metric as
\be
g_{MN} d x^M dx^N = {K^2 \over r^2} 
\left[dr dr + \hat{g}_{\mu\nu} dx^\mu dx^\nu \right], 
\label{gaugefixing}
\ee
where $\hat{g}_{\mu\nu}$ is the arbitrary five dimensional metric 
and the cosmological constant is given by 
\be
\Lambda = K^{-2} = {6 \sqrt{3}} g^{3 \over 2} m^{1 \over 2}.
\ee
We choose the gauge satisfying
\be
e_r{}^5 = {K \over r}, 
\qquad 
e_r{}^a = e_\mu{}^5 = A_r^I = \psi_{ri} = 0, 
\ee
for which the rest of the fields are arbitrary. 
Note that our choice is just an which leads to a successful result. 
Next, we calculate the $r$-dependence of the fields. 
Define the fluctuation $\varphi=\phi-\phi_0$, 
then we obtain the field equation for $\varphi$ as follows:
\be
r^6 \partial_r ( r^{-4} \partial_r \varphi) 
+ 6 \varphi 
= 0, 
\ee
where we omit the higher order terms in $r$.
Then we see that $r$-dependence of $\varphi$ is $r^2$ or $r^3$. 
It is enough to choose the dominant one $\varphi \sim r^2$, 
since we want to take the near boundary limit $r\rightarrow 0$. 
Similarly the linearized field equation for $A_\mu^I$ gives, 
$A_\mu^I \sim r^0$. 
\par
As for $B_{MN}$, consider first the $r$-dependence of 
$B^{\mu\nu}$ and $B^{\mu r}$ since we have 
${\cal D}_\mu B^{r\rho}$ in the field equations. 
The field equations read: 
\be
3{\cal D}_M {\cal D}^{[M} B^{NP]} 
- 2 K^{-2} B^{NP}= 0. 
\label{simplyanti}
\ee
Multiplying ${\cal D}_N$ to this equation, we obtain 
\be
{\cal D}_N B^{NP} = 0.
\ee
Then we substitute it back to (\ref{simplyanti}), 
and finally we get the field equation as 
\be
\{ {\cal D}_M {\cal D}^M 
- 6 K^{-2} \} B^{NP}=0. 
\ee
We have to know the order in $r$ for these terms, so that 
we write down the part of the above equation explicitly: 
\be
{\cal D}_M B^{M \nu} 
= \partial_r (\sqrt{-g} g^{rr} g^{\nu\rho} B_{r\rho}) 
+ \partial_\mu (\sqrt{-g} g^{\mu\lambda} g^{\nu\rho} B_{\lambda\rho}) , 
\label{BfieldEOM}
\ee
where $B_{r\rho}$ gives the higher order in $r$ of $B_{\lambda\rho}$, 
since these two terms in (\ref{BfieldEOM}) have the same power in $r$. Thus, 
we can treat 
${\cal D}_\mu B^{r\rho}
= -{1 \over r} \hat{g}_{\mu\nu} B^{\nu\rho} +$ higher order, 
so that we cannot neglect $B_{r\rho}$ term. 
The linearized field equations $B^{\mu\nu}$ are now 
\be
r^2 \partial_r^2 B^{\nu\rho} -8 r B^{\nu \rho} + 18 B^{\nu\rho} 
= 0. 
\ee
Thus finally we obtain the behavior of the antisymmetric tensor field as 
$B_{\mu\nu} \sim r^{-1}$. 
\par
In the case of fermionic fields, 
we introduce $\tilde{\Gamma} \equiv \Gamma^5 \Gamma_7$ 
and define the projection operators 
$P_{\pm}={1 \over 2} (1 \pm \tilde{\Gamma})$. 
They play a role of `chiral projection on operators' 
in the odd dimensional AdS supergravity. 
For even dimensional AdS supergravity, we can define 
the chirality, whereas we cannot in odd dimensions. 
We assign the positive eigenvalue of $\tilde{\gamma}$ 
to the upper component $\psi_+$, and the negative one to 
the lower component $\psi_-$. 
\par
We multiply $P_{\pm}$ to the field equations of $\chi_i$, and we have 
\be
r \Gamma^5 \partial_r \chi_{i\mp} + r \Gamma^\mu \partial_\mu \chi_{i\pm}
+{5 \over 2} \Gamma^5 \chi_{i\mp} + \Gamma_7 \chi_{i\mp} = 0. 
\label{chieom}
\ee
First, to know the main behavior, we omit the second term which 
is a higher order one. 
By multiplying $\Gamma^5$ from the left hand side, 
we obtain 
\ba
r \partial_r \chi_{i+} + {5 \over 2} \chi_{i+} + \chi_{i+} 
\A = \A 0, 
\nonu 
r \partial_r \chi_{i-} + {5 \over 2} \chi_{i-} - \chi_{i-}
\A = \A 0. 
\ea
We find that $\chi_{i\pm}$ behave as
\be
\chi_{i+} \sim r^{3 \over 2}, \qquad \chi_{i-} \sim r^{7 \over 2}. 
\ee
Then we substitute these results into (\ref{chieom}), 
we obtain two sets of solution 
$\chi_{i+} \sim r^{3 \over 2}, \quad \chi_{i-} \sim r^{5 \over 2}$ 
and 
$\chi_{i+} \sim r^{9 \over 2}, \quad \chi_{i-} \sim r^{7 \over 2}$. 
We choose the dominant behavior set
\be
\chi_{i+} \sim r^{3 \over 2}, \qquad \chi_{i-} \sim r^{5 \over 2}. 
\ee
\par
Similarly, we obtain $r$-dependence 
of the Rarita-Schwinger fields as 
$\psi_{\mu i-} \sim r^{- {1 \over 2}}$ and 
$\psi_{\mu i+} \sim r^{+ {1 \over 2}}$, respectively. 
%%%%% Section 4 %%%%%%%%%%%%%%%%%%%%%%%%%%%%%%%%%%%%%%%%%%%%
%
\newsection{Local symmetries on the boundary}
In what follows, using the supergravity fields whose 
$r$-dependence were obtained explicitly 
in the previous section 
to solve $r$-dependence of 
the local transformation parameters. 
We obtained in the last section 
the boundary behaviors of the fields as
\ba
e_\mu{}^a = \left({r \over K}\right)^{-1} 
\hat{e_\mu{}^a}, \quad 
\psi_{\mu i+} \A = \A \left({r \over K}\right)^{1 \over 2} 
\hat{\psi}_{\mu i+}, \quad 
\psi_{\mu i-} = \left({r \over K}\right)^{-{1 \over 2}} 
\hat{\psi}_{\mu i-}, \nonu
B_{\mu\nu} = \left({r \over K}\right)^{-1} 
\hat{B}_{\mu\nu}, \quad 
B_{\mu r} \A = \A \left({r \over K}\right)^{0} 
\hat{B}_{\mu r}, \quad 
A_\mu^I = \left({r \over K}\right)^{0} 
\hat{A}_\mu^I, 
\nonu
\chi_{i+} = \left({r \over K}\right)^{3 \over 2} 
\hat{\chi}_{i+}, \quad
\chi_{i-} \A = \A \left({r \over K}\right)^{5 \over 2} 
\hat{\chi}_{i-}, \quad
\varphi = \left({r \over K}\right)^2 \hat{\varphi}. 
\label{fields}
\ea
Here the hatted fields become arbitrary functions of $x^\mu$, 
being independent of $r$ on the boundary. 
\par
We substitute (\ref{fields})  and 
the gauge fixing conditions (\ref{gaugefixing}) into 
the transformation law in six dimensions. 
To keep the gauge fixing conditions, 
we need some constraints on the parameters, 
whose solution represents the `residual symmetry'. 
For example, from the transformation law for the $e_r{}^5$, we read 
\be
\delta \hat{e}_r{}^5 = \xi^r \partial_r \hat{e}_r{}^5 
+ \xi^\nu \partial_\nu \hat{e}_r{}^5
+ \partial_r \xi^r \hat{e}_r{}^5 
+ \partial_r \xi^\nu \hat{e}_\nu{}^5
+ \Sigma^5{}_a \hat{e}_r{}^a 
+ \bar{\hat{\psi}}^i_r \gamma^5 \epsilon_i, 
\ee
which should vanish because of the gauge fixing. Then we have
\be
\partial_r \left({{1 \over r} \xi^r}\right)=0, 
\ee
and its solution is $\xi^r \sim r$. 
We can finally obtain $r$-dependence for the transformation 
parameters as
\ba
\epsilon^{\pm} 
= \left({r \over K}\right)^{\pm{1 \over 2}}
\hat{\epsilon}^{\pm} , 
\quad 
\xi^r  
\A = \A \left({r \over K}\right)^1 \hat{\xi}^r, 
\quad 
\xi^\nu  
= \left({r \over K}\right)^0 \hat{\xi}^\nu,  
\nonu
\Lambda^I 
= \left({r \over K}\right)^2 \hat{\Lambda}^I, 
\quad 
\Sigma^{ab}
\A = \A \left({r \over K}\right)^0 \hat{\Sigma}^{ab},
\quad 
\Sigma^{a5}
= \left({r \over K}\right)^1 \hat{\Sigma}^{a5}, 
\label{params}
\ea
where the hatted functions mean the one of the boundary coordinate 
$x^\mu$ and do not depend on the $r$-direction, as before. 
\par
Before substituting (\ref{fields}) and (\ref{params}) 
into the local transformations of the six dimensional 
AdS supergravity (\ref{transfs6}), 
we have to examine the fields which are not independent, 
and express them by the other fields. 
To do this, 
we use the equations of motion for the fields including 
the interaction terms but ignoring 2-fermi terms, 
since they contribute to the higher order fermi terms. 
Then we pull out the dominant contributions 
having the lowest power of $r$ by taking the near boundary limit, 
and finally express the field in terms of the independent fields. 
As for the fermion fields, 
we deal with the fields having the higher order 
powers in $r$. 
Now we express the fields which is obtained by taking the 
$r \rightarrow 0$ limit as the ones with suffix 0. 
We have
\be
B_{0\mu r} = {1 \over \sqrt{2}m} 
\left( {3m \over g} \right)^{3 \over 4} 
{\cal D}_{0 \nu} B_0^\nu{}_\mu
+{1 \over 4}
\left( {3m \over g} \right)^{1 \over 2} 
e_0^{-1} \epsilon_{\mu\nu\rho\sigma\tau} 
B_0^{\nu\rho}B_0^{\sigma\tau}.
\label{antisym} 
\ee
\par
As for $\psi_{0\mu i-}$, we apply the projection operator $P_-$ 
onto the field equation. Then we find the lowest order is 
${\cal O}(r^{5 \over 2})$, and we obtain 
\be
-{3 \over K} \Gamma_0^{\mu\nu} \Gamma^5 \psi_{0\nu i+} 
= \Gamma_0^{\mu\nu\rho} {\cal D}_{0\nu} \psi_{0\rho i-}
-{3m \over 4 \sqrt{2}} \left( {3m \over g}\right)^{-{1 \over 4}}
B_0^{\rho\sigma}\Gamma_0^{[\mu}\Gamma_{0\sigma\rho}\Gamma_0^{\nu]}
\psi_{0\nu i-}. \label{rseom}
\ee
Let us define $\phi^\mu_{0i-}$ by 
\be
\phi^\mu_{0i-} \equiv 
\Gamma_0^{\mu\nu\rho} {\cal D}_{0\nu} \psi_{0\rho i-}
-{3m \over 4 \sqrt{2}} \left( {3m \over g}\right)^{-{1 \over 4}}
B_0^{\rho\sigma}\Gamma_0^{[\mu}\Gamma_{0\sigma\rho}\Gamma_0^{\nu]}
\psi_{0\nu i-}. 
\ee 
Then, solving (\ref{rseom}) we can express $\psi_{0\mu i+}$ 
only in terms of independent fields, namely, 
\be
\psi_{0\mu i+} = - {K \over 3} \Gamma^5
\left( 
g_{0\mu\nu}-{1 \over 4}\Gamma_{0\mu} \Gamma_{0\nu}
\right)
\phi^\nu_{i-}. 
\label{raritaschwinger}
\ee
\par
For $\chi_{0i+}$, 
we use the projection $P_+$, 
and from the lowest order terms of 
${\cal O}(r^{5 \over 2})$, we can obtain 
\ba
-{1 \over K} \Gamma_7 \chi_{0i-} 
\A = \A \Gamma_0^\mu {\cal D}_{0\mu} \chi_{0i+}
-{m \over 2\sqrt{2}} \left( {3m \over g}\right)^{-{1 \over 4}} 
\underline{B_{0\mu r}} 
\Gamma_0^\rho \Gamma_0^\mu \Gamma^5 \psi_{0\rho i-} 
\nonu 
\A - \A {m \over 2 \sqrt{2}} 
\left( {3m \over g}\right)^{-{1 \over 4}} 
B_{0\mu\nu}\Gamma_0^\rho \Gamma_0^{\mu\nu} 
\underline{\psi_{0\rho i+}}
\nonu 
\A + \A {1 \over 4 \sqrt{2}}i 
\left( {3m \over g}\right)^{-{1 \over 4}} 
F_{0\mu\nu}\Gamma_0^\rho \Gamma_0^{\mu\nu} \Gamma_7 
(\sigma^I)_i{}^j
\psi_{0\rho j-}
- {m \over 2 \sqrt{2}} 
\left( {3m \over g}\right)^{-{1 \over 4}} 
B_{0\mu\nu} \Gamma_0^{\mu\nu} \chi_{0i+}
\nonu
\A - \A {1 \over 4}
\left( {3m \over g}\right)^{-{1 \over 4}} 
\partial_\mu B_{0\nu\rho}
\Gamma_0^\sigma \Gamma_0^{\mu\nu\rho}\Gamma_7 \psi_{0\sigma i-}
+ 
{1 \over \sqrt{2}K} \varphi_0 \Gamma_0^\mu \Gamma^5 \psi_{0\mu i-}.
\label{spinor}
\ea
The underlined fields are expressed by (\ref{antisym}) 
and (\ref{raritaschwinger}). 
\par
Furthermore, we decompose 16 component spinors $\psi$
into two eight component ones, 
since in five dimensions, 
the number of components of a spinor is eight.
We decompose $\psi$ as
\ba
\psi_\pm \A = \A {1 \over 2} (1 \pm \tilde{\Gamma}) \psi, \nonu
\psi_- = \left(
      \begin{array}{c}
      0 \\
      \psi^{\rm D} 
      \end{array}
      \right) \A , \A 
      \psi_+ = \left(
      \begin{array}{c}
      \psi^{\rm U}\\
      0
      \end{array}
      \right) .
\ea
Here, we have represented the gamma matrices as 
\ba
\Gamma^a \A = \A \gamma^a \otimes \sigma^3 ,\nonu
\Gamma^5 \A = \A {\bf 1} \otimes \sigma^1 ,\\
\Gamma^7 \A = \A {\bf 1} \otimes (-i \sigma^2) , \nonumber
\ea
so that $\tilde{\Gamma}={\bf 1} \otimes \sigma^3$. 
\par
We redefine some of the fields to simplify the equations.
For the antisymmetric tensor $B_{\mu\nu}$, 
to eliminate the derivative of $\epsilon_{0i-}$ in 
the supertransformation, we define 
\be
\tilde{B}_{0\mu\nu} \equiv B_{0\mu\nu} + {1 \over \sqrt{2}m} 
\left(
{3m \over g}
\right)^{1 \over 4}
\bar{\psi}_{0[\mu}^i\psi_{0\nu] i} .
\ee
We also rescale the fields as
\ba
\psi_{0\mu i+} \rightarrow K \psi_{0\mu i+} , \quad 
B_{0\mu\nu} \rightarrow  
{\sqrt{2} \over m} 
\left(
{3m \over g}
\right)^{1 \over 4}
B_{0\mu\nu} , \quad 
B_{0\mu r} \rightarrow {3 \over mg}B_{0\mu r}
\nonu 
A_{0\mu}^I \rightarrow {1 \over g} A_{0\mu}^I , \quad
\chi_{0i+} \rightarrow K \chi_{0i+}, \quad 
\chi_{0i-} \rightarrow K^2 \chi_{0i-} ,\quad
\varphi_0 \rightarrow K^2 \varphi_0 ,
\ea
such that the transformations become independent 
of $m$ and $g$.
Also let define $\epsilon_{0i+} = -K\Gamma_7 \eta_{0i-}$, 
where the $\eta_{0i-}$ is an arbitrary function of $x^\mu$.
\par
Then, the definitions of various quantities become 
\ba
D_{0\mu}\epsilon_{0i}^{\rm D} 
\A = \A \left(
     \partial_\mu + {1 \over 4} \omega_{0\mu}{}^{ab}\gamma_{ab}
     \right) \epsilon_{0i}^{\rm D} 
     -i{1 \over 2}A_{0\mu}^I(\sigma^I)_i{}^j
     \epsilon_{0j}^{{\rm D}} ,
\nonu 
F_{0\mu\nu}^I \A = \A 2 \partial_{[\mu}A_{0\nu]}^I
     +\epsilon^{IJK} A_{0\mu}^J A_{0\nu}^K ,
\nonu 
\tilde{B}_{0\mu\nu} \A = \A B_{0\mu\nu}
     -{1 \over 2}\bar{\psi}_{0[\mu}^{{\rm D}i} \psi_{0\nu]i}^{\rm D}, 
\nonu
\phi_{0i}^{{\rm D}\mu}
\A = \A -i 
      \gamma^{\mu\nu\rho} {\cal D}_{0\nu} \psi_{0\rho i}^{\rm D}
      -{3 \over 4}B_0^{\rho\sigma} 
      \gamma^{[\mu}\gamma_{\rho\sigma}\gamma^{\nu]}
      \psi_{0\rho i}^{\rm D}.
\ea
The expressions of the dependent fields become
\ba
\psi_{0\mu i}^{\rm U} \A = \A 
     -i{1 \over 3}\left(
     g_{0\mu\nu}-{1 \over 4}\gamma_\mu\gamma_\nu
     \right) \phi_{0i}^{{\rm D}\nu}, 
\nonu
B_{0\mu r}\A =\A 
     {\cal D}_{0\nu}\tilde{B}^\nu_{0\mu}
     +{1 \over 2} e^{\mu\nu\rho\sigma\tau} 
     \tilde{B}_{0\nu\rho} \tilde{B}_{0\sigma\tau},
\nonu 
\chi_{0i}^{\rm D} \A = \A 
     -i \gamma^\mu {\cal D}_{0\mu} \chi_{0i}^{\rm U}
     -i{1 \over 4} \underline{B_{0\mu r}}
          \gamma^\rho\gamma^\mu \psi_{0\rho i}^{\rm D}
     -i{1 \over 2}\tilde{B}_{0\mu\nu}\gamma^\rho\gamma^{\mu\nu}
          \psi_{0\rho i}^{\rm D}
          \nonu \A \A 
     +{1 \over 24}F_{0\mu\nu}\gamma^{\mu\nu}\chi_{0i}^{\rm U}
     +{1 \over 4}\partial_\mu \tilde{B}_{0\nu\rho}
          \gamma^\sigma\gamma^{\mu\nu\rho}
          \psi_{0\sigma i}^{\rm D}
     -{1 \over \sqrt{2}}\varphi_0 \gamma^\mu 
          \psi_{0\mu i}^{\rm D}.
\label{indepfields}
\ea
The underlined part in the third expression is written by the second one.
\par
Taking $r \rightarrow 0$ limit, we obtain 
the local transformations on the boundary
\ba
\delta e_{0\mu}{}^a \A = \A \xi_0^\nu \partial_{0\nu} e_{0\mu}{}^a 
+ \partial_{0\mu} \xi_0^\nu e_{0\nu}{}^a 
+ \Omega_0 e_{0\mu}{}^a 
+ \Sigma^a_{0b} e_{0\mu}{}^b
- i \bar{\psi}_{0\mu}^{{\rm D} i} \gamma^a \epsilon_{0i}^{\rm D}, 
\nonu
\delta A_{0\mu}^I \A = \A \xi_0^\nu \partial_{0\nu} A_{0\mu}^I 
+ \partial_{0\mu} \xi_0^\nu A_{0\nu}^I 
+ {\cal D}_{0\mu} \Lambda_0^I 
\nonu \A \A 
-{3 \over 2} \bar{\chi}_0^{i \rm U} (\sigma^I)_i{}^j 
     \epsilon_{0j}^{\rm D}
-3 i \underline{\bar{\psi}_{0 \mu}^{{\rm U} i}} 
     (\sigma^I)_i{}^j \epsilon_{0j}^{\rm D}
-3 i \bar{\psi}^{i \rm D}_{0\mu}(\sigma^I)_i{}^j \eta_{0j}^{\rm D}, 
\nonu
\delta \varphi_0 \A = \A 
     \xi_0^\mu \partial_\mu \varphi_0 
          -2 \Lambda_0 \varphi_0 
     + {1 \over \sqrt{2}} 
          (- \underline{\bar{\chi}_0^{{\rm D}i}} \epsilon_{0i}^{\rm D}
     + \bar{\chi}_0^{{\rm U}i} \eta_{0i}^{\rm D}), 
\nonu
\delta \tilde{B}_{0\mu\nu} \A = \A 
\xi_0^\rho \partial_\rho \tilde{B}_{0\mu\nu} 
+ \partial_\mu \xi_0^\rho \tilde{B}_{0\rho\nu} 
+ \partial_\nu \xi^\rho \tilde{B}_{0\mu\rho} 
+ \Omega_0 \tilde{B}_{0\mu\nu} 
\nonu \A \A 
      + {1 \over 2}\bar{\chi}_0^{{\rm U}i} 
           \gamma_{\mu\nu} \epsilon_{0i}^{\rm D}
      + {\cal D}_{0[\mu}\gamma_{\nu]} \epsilon_{0i}^{\rm D}
      + \underline{\bar{\psi}_{0[\mu}^{{\rm U}i}}
           \gamma_{\nu]} \epsilon_{0i}^{\rm D}
      + i{1 \over 4} \bar{\psi}_{0[\nu}^{{\rm D}i}
           (\gamma_{\mu]}{}^{\rho\sigma}-4\delta_{\mu]}^\rho \gamma^\sigma)
           \epsilon_{0i}^{\rm D}\tilde{B}_{0\rho\sigma}, 
      \nonu
\delta \psi_{0\mu i}^{\rm D} 
     \A = \A \xi_0^\nu \partial_\nu \psi_{0\mu i}^{\rm D} 
          + \partial_\mu \xi_0^\nu \psi_{0\nu i}^{\rm D} 
          +{1 \over 2} \Omega_0 \psi_{0\mu i}^{\rm D} 
          + {1 \over 4} \Sigma^{0ab} \gamma_{ab} \psi^{{\rm D} i}_{0\mu }
          -i {1 \over 2} g \Lambda_0^I (\sigma^I)_i{}^j \psi^{\rm D}_{0\mu j}
     \nonu \A \A 
     + {\cal D}_{0\mu} \epsilon_{0i}^{\rm D}
     -i {1 \over 4} \tilde{B}_{0\nu\rho}
          (\gamma_\mu{}^{\nu \rho}-4 \delta_\mu^\nu \gamma^\rho)
          \epsilon_{0i}^{\rm D}
     - i \gamma_\mu \eta_{0i}^{\rm D}, 
\nonu 
\delta \chi_{0i}^{\rm U} \A = \A 
\xi_0^\mu \partial_\mu \chi_{0i}^{\rm U}
-{3 \over 2}\Omega_0 \chi_{0i}^{\rm U}
-{1 \over 4}\Sigma_0^{ab}\gamma_{ab} \chi_{0i}^{\rm U}
-i {1 \over 2} g \Lambda_0^I (\sigma^I)_i{}^j \chi^{\rm U}_{0j}
\nonu \A \A 
     -{1 \over \sqrt{2}} \varphi_0 \epsilon_0^{{\rm D}i} 
-i {1 \over 4} \gamma^{\mu\nu\rho}\partial_{[\mu} \tilde{B}_{0\nu\rho]}
     \epsilon_{0i}^{\rm D}
-i{1 \over 24} F_{0\mu\nu}^I \gamma^{\mu\nu} 
     (\sigma^I)_i{}^j \epsilon_j^{\rm D}
\nonu \A \A 
+ {1 \over 4} \underline{B_{0r\nu}} \epsilon_{0i}^{\rm D}
-{1 \over 2} \tilde{B}_{0\mu\nu} \gamma^{\mu\nu}
     \eta_{0j}^{\rm D}. 
\ea
{}From these equations, 
we find $\Omega_0$ is the Weyl transformation parameter, 
and $\eta^{\rm D}_{0i}$ that of the super Weyl transformation. 
The underlined terms are expressed by (\ref{indepfields}). 
We can count the off-shell degrees of freedom of the above fields: 
\ba
\rm{d.o.f}(e_{0\mu}{}^a)=9, \quad \A \A 
\rm{d.o.f}(A_{0\mu}^I)=12, \quad 
\rm{d.o.f}(\varphi_0)=1, \quad
\rm{d.o.f}(\tilde{B}_{0\mu\nu})=10,
\nonu 
\A \A 
\rm{d.o.f}(\psi_{0\mu i}^{\rm D})=24, \quad
\rm{d.o.f}(\chi_{0i}^{\rm U})=8. 
\ea
We see that the bosonic degrees of freedom 
and the fermionic one 
are the same. 
\par
%
%%%%% Section 5 %%%%%%%%%%%%%%%%%%%%%%%%%%%%%%%%%%%%%%%%%%%%
%
\newsection{Summary and Discussions}
We have attempted to construct a certain five-dimensional CSG. 
As a first step of this attempt, 
the local transformation laws of this CSG are determined.
In general, 
the coefficients of 
the Weyl transformation parameter $\Omega_0$ 
should be the Weyl weights, so that we could predict them 
in the five dimensional conformal supergravity. 
\par
A few interesting problems have remained. 
Although it is rather tedious work, we can 
construct the CSG in the standard fashion 
\cite{KTvN}, \cite{vN}, \cite{SS}, 
and compare the result with that obtained in this paper.  
The construction would be very similar to that of the 
six dimensional CSG \cite{BSvP} since the gauge group 
in both theories is the same SU(2). 
\par
Furthermore, 
our result would give a key to understand 
the field theory of D4-D8 system 
\cite{SEIBERG}, \cite{BO}, \cite{FKPZ}. 
Actually, our CSG field contents in five dimensions 
represent the pure gravity parts of the fields in 
the result of D4-D8 system \cite{FKPZ}. 
\par
If we execute the same method \cite{NT2} using these CSG 
transformations, we will get the eta invariants instead of the anomalies 
of local symmetries\footnote{The author would like to thank E. Kiritsis 
for pointing it out.}. 
%
%
%%%%%  Acknowledgements  %%%%%%%%%%%%%%%%%%%%%%%%%%%%%%%%%
%
\begin{flushleft}
\large{\bf{Acknowledgements}}
\end{flushleft}
The author would like to thank Y. Tanii for useful discussions 
on the supergravities and A. Sugamoto for his careful reading of the 
manuscript. She would also like to thank the members of Stockholm 
University, especially A. Fayyazuddin, and the Niels Bohr Institute 
for their warm hospitality. 
%
%
%
%%
%%%%%  References  %%%%%%%%%%%%%%%%%%%%%%%%%%%%%%%%%%%%%%%
%
\newcommand{\NP}[1]{{\it Nucl.\ Phys.\ }{\bf #1}}
\newcommand{\PL}[1]{{\it Phys.\ Lett.\ }{\bf #1}}
\newcommand{\CMP}[1]{{\it Commun.\ Math.\ Phys.\ }{\bf #1}}
\newcommand{\MPL}[1]{{\it Mod.\ Phys.\ Lett.\ }{\bf #1}}
\newcommand{\IJMP}[1]{{\it Int.\ J. Mod.\ Phys.\ }{\bf #1}}
\newcommand{\PRe}[1]{{\it Phys.\ Rev.\ }{\bf #1}}
\newcommand{\PRL}[1]{{\it Phys.\ Rev.\ Lett.\ }{\bf #1}}
\newcommand{\PTP}[1]{{\it Prog.\ Theor.\ Phys.\ }{\bf #1}}
\newcommand{\PTPS}[1]{{\it Prog.\ Theor.\ Phys.\ Suppl.\ }{\bf #1}}
\newcommand{\APe}[1]{{\it Ann.\ Phys.\ }{\bf #1}}
\newcommand{\ATMP}[1]{{\it Adv.\ Theor.\ Math.\ Phys.\ }{\bf #1}}
\end{document}